\documentstyle[aps,preprint,prl,psfig,epsfig,amssymb]{revtex}

\tighten
\begin{document}
\bibliographystyle{prsty}

\title{Finite-size scaling at the dynamical transition of the 
       mean-field 10-state Potts glass}
\author{Claudio Brangian, Walter Kob, and Kurt Binder}
\address{Institute of Physics, Johannes Gutenberg University,
        Staudinger Weg 7, D55099 Mainz, Germany}
\maketitle

\begin{abstract}
We use Monte Carlo simulations to study the static and dynamical
properties of a Potts glass with infinite range Gaussian distributed
exchange interactions for a broad range of temperature and system
size up to $N=2560$ spins. The results are compatible with a critical
divergence of the relaxation time $\tau$ at the theoretically
predicted dynamical transition temperature $T_D$, $\tau \propto
(T-T_D)^{-\Delta}$ with $\Delta \approx 2$. For finite $N$ a further
power law at $T=T_D$ is found, $\tau(T=T_D) \propto N^{z^\star}$ with
$z^\star \approx 1.5$ and for $T>T_D$ dynamical finite-size scaling
seems to hold.  The order parameter distribution $P(q)$ is
qualitatively compatible with the scenario of a first order glass
transition as predicted from one-step replica symmetry breaking
schemes.
\end{abstract}

\begin{center}
29. September, 2000
\end{center}

Developing a theory of the glass transition of a
fluid from its Hamiltonian within first principles
statistical mechanics is still a formidable problem
~\cite{Jackle:1986,Gotze:1989,Parisi:1997,Mezard:1999}. While some
researchers attribute glassy freezing to the (hypothetical) vanishing
of the configurational entropy ~\cite{Gibbs:1958} at the ``Kauzmann
temperature'' $T_K$ ~\cite{Kauzmann:1948} (which is lower than the
experimental ~\cite{Jackle:1986} glass transition temperature $T_g$),
others emphasize the dynamical transition at the critical temperature
$T_c$ of mode coupling-theory ~\cite{Gotze:1989} from the ergodic fluid
to a non ergodic state. Since, for atomic systems, $T_c > T_g$, this
frozen phase can have only a finite lifetime and is thought to decay by
thermally activated (so-called ``hopping'') processes.

Recently evidence has been given~\cite{Parisi:1997,Mezard:1999} that these
two seemingly different scenarios could both result as two complementary
aspects of the same unifying theory ~\cite{Stillinger:1988}. In
view of the questions that still exist on the various theoretical
approaches, it is valuable to have exactly solvable models that exhibit
a similar behavior: a dynamical transition at a temperature $T_D$ and
a static first order glass transition at a temperature $T_0 < T_D$. One
of these models is the $p$-state infinite range Potts glass with $p>4$
~\cite{Gross:1985,Elderfield:1983,Cwilich:1989,DeSantis:1995,Kirkpatrick:1987,Kirkpatrick:1988},
where at $T_0$ a static (Edwards-Anderson type
~\cite{Edwards:1975,Binder:1986}) spin glass order parameter
$q_{EA}$ appears discontinuously. However at $T_0$ there is neither a
latent heat nor a divergence of the static spin glass susceptibility
$\chi_{SG}$. The latter would diverge only at an extrapolated spinodal
temperature $T_s<T_0$, see Fig. 1.  The dynamical behavior of the spin
autocorrelation function $C(t)$ for $T \gtrsim T_D$ is described by the
same type of equations ~\cite{Kirkpatrick:1987,Kirkpatrick:1988} as found
in mode-coupling theory ~\cite{Gotze:1989}. Thus this model seems indeed
to have many properties in common with structural glasses. Apart from
being a possible prototype model for the structural glass transition,
the Potts glass can also be considered as a simplified model for an
anisotropic orientational glass: e.g. a six-state Potts glass may be
a reasonable description of a diluted cubic molecular crystal where
diatomic molecules can align only along the $p=6$ face diagonals
~\cite{Binder:1998}.

In order to understand this model in more detail, we have performed
extensive Monte Carlo simulations. Of course these simulations were
done for systems with finite size and thus all the transitions in
Fig. 1 are rounded. While in the past finite size effects at normal
first and second order transitions have been studied extensively
~\cite{Privman:1990,Binder:1992}, very little is known about finite
size effects at \textsl{dynamical} transitions. Thus it is hoped that
the present work will be useful for the proper analysis of simulations
of realistic models for the structural glass transition as well
~\cite{Kob:1999}. In addition, \textsl{finite} mean-field systems may 
also have some similarities with systems having finite interaction range
~\cite{Crisanti:2000}: the relaxation time $\tau$ is large but finite
for $T < T_D$ as well, unlike the behavior in the thermodynamic limit
$N\rightarrow\infty$ (shown in Fig. 1), since energy barriers between
``basins'' are finite for $T<T_D$ ~\cite {Crisanti:2000}. Thus the
mean field Potts glass for finite $N$ should be a good model for
elucidating the physics of glassy systems in general.

We now give some technical details about our simulations. We study the
Hamiltonian
\begin{equation}
H=-{1 \over 2}\sum_{i\neq j}^{N}J_{ij}(p\delta_{\sigma_i \sigma_j}-1),
\end{equation}
where each spin $\sigma_i \in \{1,\ldots,p\}$ interacts with all the
other spins. The interactions $J_{ij}$ are taken from a Gaussian
distribution with mean $J_0=(3-p)/(N-1)$ and variance $\Delta J =
(N-1)^{-1/2}$. We choose $p=10$ (note that for $p=2$ the standard Ising
spin glass results ~\cite{Binder:1986}). For this choice of $\Delta J$ it
can be shown~\cite{Elderfield:1983} that $T_s=1$ ~\cite{second_transition}.
We simulate the system sizes $N$=160, 320, 640, 1280 and 2560, and
the number of independent samples is 500 for $N$=160 and $T>1$, else
100, with the exception of $N$=2560 where only 50 samples where used.
The equilibrium dynamics is studied by means of the standard Metropolis
algorithm ~\cite{Binder:1986,Binder:1998,Privman:1990}.  In order to
produce equilibrated configurations and to study the static properties
we used again the Metropolis algorithm at high temperature ($T \leqslant
T_D$), while for lower temperatures we used the parallel tempering method
~\cite{Hukushima:1996,Kob:2000}.

Fig. 2 shows typical data for the autocorrelation function $C(t)$ of
the Potts spins ~\cite{Wu:1982} at two temperatures and various system
sizes.  While at very high temperature, such as $T=1.8$, finite size
effects are completely negligible, at $T=T_D$ they are unexpectedly
pronounced. Whereas in the thermodynamic limit one expects at the
dynamical transition $C(t\rightarrow\infty)=q_{EA}$, for
$N\leqslant1280$ hardly any indication of the development of the
plateau (as expected from Fig. 1) is seen. This behavior is strikingly
different from the behavior found in analogous autocorrelation
functions for atomistic models of structural glasses, where a pronounced
plateau occurs already at temperatures a few percent above $T_c$
($=T_D$ in our model) ~\cite{Kob:1999} and finite size effects are hardly
detectable for a comparable range of $N$. While for
$N\rightarrow\infty$ the dynamics of the present model for $T\geqslant
T_D$ is described by mode-coupling theory
~\cite{Kirkpatrick:1987,Kirkpatrick:1988}, the assertion that infinite
range models for finite $N$ resemble the behavior of real structural
glasses ~\cite{Crisanti:2000} seems, in view of Fig. 2, doubtful to
us, at least for the system sizes presently accessible.

In order to quantify the slowing down of the dynamics as a function
of the temperature and system size we define a relaxation time
$\tau$ by $C(\tau)=0.2$. This particular choice was made since we
must have $C(\tau)<q_{EA}(T=T_D)$. Our results indicate that, for
$T=T_D $, $\tau$ behaves like a power law $\tau \propto N^{z^*}$
with $z^*\approx1.5$ (see inset in Fig. 3). It is known from the
analytical results \cite{Kirkpatrick:1988} that, in the limit
$N\rightarrow\infty$, $\tau (T) \propto
(T/T_D-1)^{-\Delta}$. The dynamical finite size scaling hypothesis
~\cite{Privman:1990,Binder:1992,Hohenberg:1977} assumes a generalized
behavior for $\tau$ as a function of $N$ and $T$ for $T\gtrsim T_D$ ,
\begin{equation}
\tau=N^{z^\star}\tilde{\tau}\{N(T/T_D-1)^{\Delta/z^{\star}}\}
 \; \textrm{ for} \;
N\rightarrow\infty \; \textrm{and} \; ({T/T_D}-1)\rightarrow 0 
\label{dffs}
\end{equation}
with the scaling function $\tilde{\tau} (\xi\rightarrow\infty)
\propto\xi^{-z^*}$ to recover the proper thermodynamic limit. As can
be seen from our data in Fig. 3, this \textsl{ansatz} is satisfied
by the Potts glass in the vicinity of the dynamical transition. Our
choice of $\Delta/z^*\approx 1.3$ in Fig. 3 implies $\Delta \approx
2$, which is similar to values found for atomistic models of the
glass transition ~\cite{Kob:1999} and compatible with a direct
extrapolation $N\rightarrow\infty$ of the data for $\tau(T,N)$ for
$T>T_D$~\cite{Brangian:2001}.

We mention that Eq. (\ref{dffs}) has a well-based theoretical foundation
for second order phase transitions ~\cite{Binder:1986,Hohenberg:1977},
where in Fig. 1 $T_0=T_D=T_s$ would coincide. However, for a
dynamical transition which has no associated diverging static
susceptibility this relation is purely phenomenological. For second
order mean-field spin glass transitions one has Eq. (\ref{dffs})
with $\Delta/z^*=2\beta_{MF}+\gamma_{MF}=3$, where $\beta_{MF}=1,
\gamma_{MF}=1$ are the static mean-field exponents of the
spin glass order parameter and the susceptibility, respectively
~\cite{Dillmann:1998}. Since $\Delta=2$ ~\cite{Binder:1986} one has
$z^*=2/3$, consistent with expectation ~\cite{z_comment} and simulations
~\cite{Bhatt:1992}. Obviously, in view of the systematic deviations from
scaling that are still visible in Fig. 3, our estimates for $\Delta$
and $z^*$ clearly are somewhat tentative only.

We conclude by discussing some results about some {\it static}
properties of this model.  As usual in numerical simulations of spin
glasses, the order parameter is a measure of the overlap between
microscopic states visited by two different replicas (same realization
of disorder, but different dynamics) of the system. For Ising spin
glasses, this is just the number of spins in the same state divided by
the total number of spins.  In the Potts glass, as well as in vector
and quadrupolar spin glasses, the overlap between spins belonging to
different replicas is a tensorial quantity. Therefore we define an
overlap order parameter which is invariant under global rotations of
the spins ~\cite{Binder:1986,Binder:1998}:
\begin{equation}
q_{\alpha \beta}=\sqrt{\sum_{\mu,\nu=1}^{p-1}(q^{\mu\nu})^2} \quad 
\textrm{with} \quad
q^{\mu\nu}={1\over N}\sum_i^NS_{i,\alpha}^{\mu}S_{i,\beta}^{\nu},
\end{equation}     
where $\alpha$ and $\beta$ are the replica indices and
$S_{i,\alpha}^{\mu}$ are the components of the spins in the simplex
representation ~\cite{Wu:1982}.  The static spin glass susceptibility
is defined as $\chi_{SG}=N[\langle q_{\alpha \beta}^2 \rangle]$.  
Our Monte Carlo
results (Fig. 4) are indeed compatible with the expected behavior for
static quantities: the spin glass susceptibility remains finite at
$T=T_D$ and the order parameter distribution $P(q)$ develops a two
peak structure not by splitting off a single peak when the temperature
is lowered, as would be common for second-order transitions
~\cite{Binder:1986,Binder:1998,Privman:1990,Binder:1992}. Instead the
second peak grows near $q=q_{EA}$ continuously gaining weight for low
temperatures. This temperature dependence is consistent with a
one-step replica symmetry-breaking scenario
~\cite{Gross:1985,Kirkpatrick:1987}.

In summary we have shown that the finite size rounding of the
dynamical transition in the $p=10$ infinite range Potts glass is
compatible with a finite size scaling hypothesis, and that the
relaxation time at the dynamical transition scales like a power law,
$\tau\propto N^{z^*}$, with $z^*\approx 1.5$. The static spin glass
susceptibility converges to a finite results at $T_D$, as expected. In
contrast to atomistic models for the glass transitions, this model
allows well equilibrated simulations at $T=T_D$ and also much lower
temperatures (Fig. 4), at least for $N$ in the range of a few
hundreds.  While atomistic models allow to study questions such as the
dynamical heterogeneity only at $T>T_D$ (which would not be a relevant
temperature region in this context for experiments), the present model
allows to determine the relaxation behavior of various quantities over
a much wider temperature range. Hence from the present model we expect
that a rather complete picture of the glass transition can be
obtained, which thus should give stimulating insights into the proper
analysis of simulations for more realistic models too, and help in
clarifying the fundamental questions that still remain about the
nature of glassy freezing in structural glasses.

\vskip 0.5 cm
\noindent
Acknowledgments: One of us (C.B.) was partially supported by the
Deutsche Forschungsgemeinschaft, Sonderforschungsbereich 262/D1. We
are grateful to the John von Neumann Institute for Computing (NIC
J\"ulich) for a generous grant of computer time at the CRAY-T3E.

\newpage
\noindent

\begin{figure}[h]
\epsfig{figure=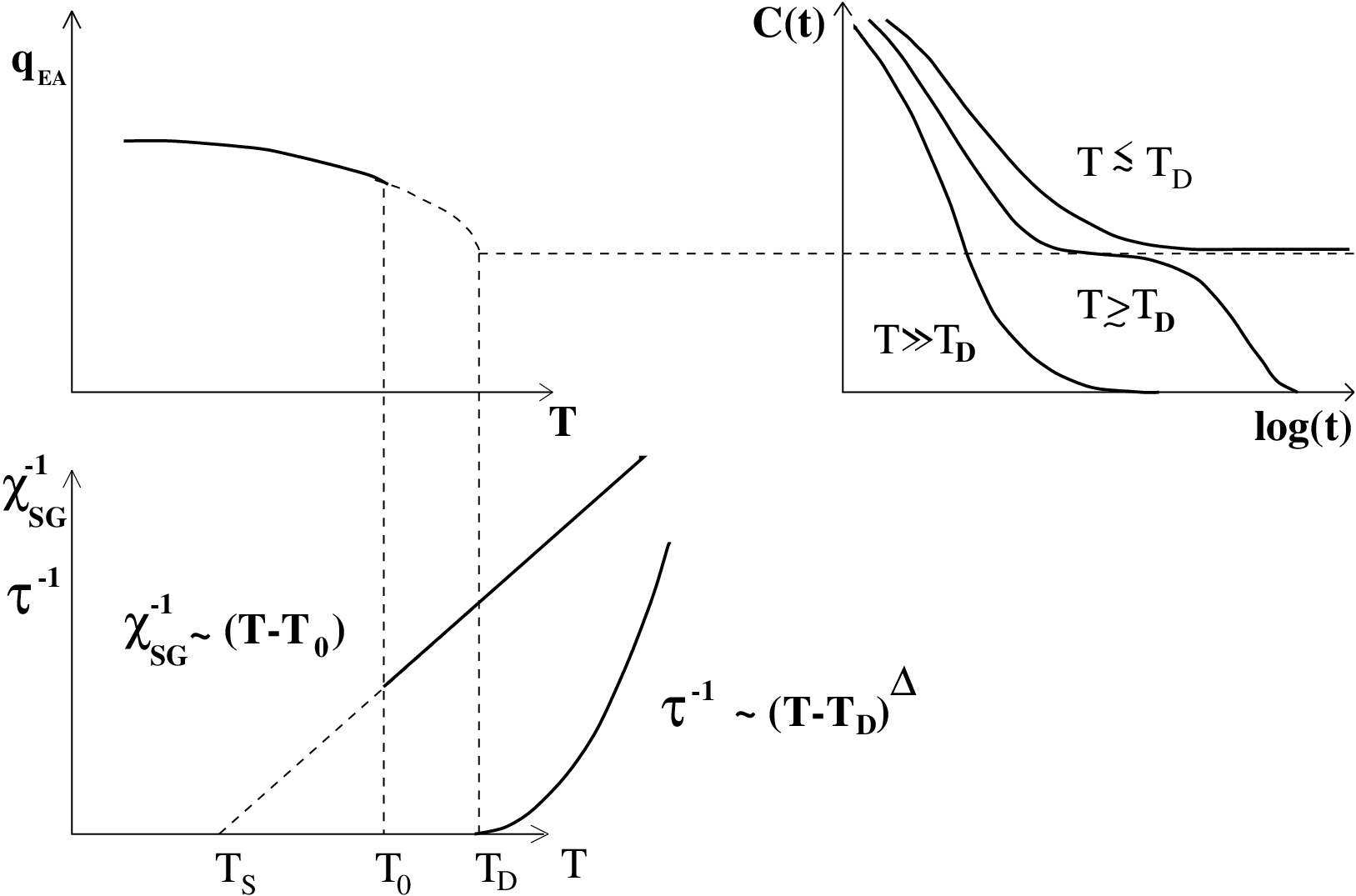,width=7.9cm,height=10.0cm, angle=270}
\caption{ Qualitative sketch of the mean-field predictions for the 
$p-$state Potts glass model with $p>4$. The spin glass order parameter in
thermal equilibrium is positive only for $T<T_0$ and jumps to zero
discontinuously at $T=T_0$, where the spin glass susceptibility
$\chi_{SG}$ is finite (for $T>T_0,\; \chi_{SG}$ follows a
Curie-Weiss type relation with an apparent divergence at
$T_s<T_0$). The relaxation time diverges already at the dynamical
transition temperature $T_D$. This divergence is due to the occurrence
of a long lived plateau in the time-dependent spin autocorrelation
function $C(t)$.}
\end{figure}

\begin{figure}[h]
\psfig{figure=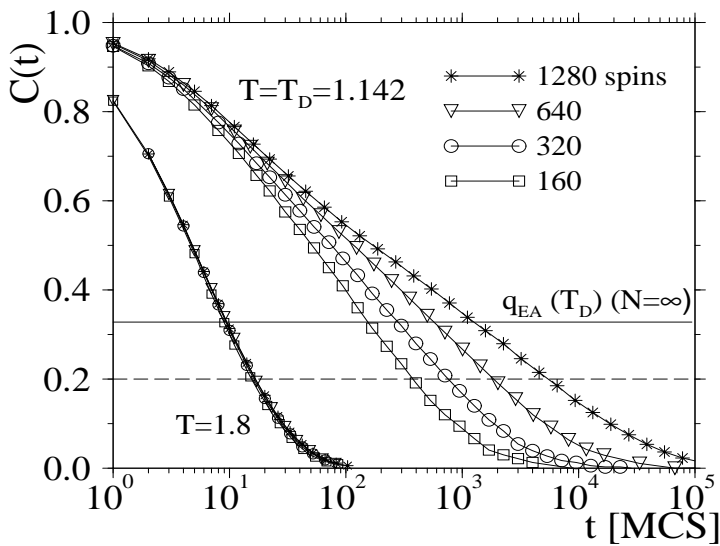,width=9.5cm,height=7.5cm}
\caption{ Spin-spin autocorrelation function $C(t)$ for $T=1.8$ and for
$T=T_D=1.142$~\protect\cite{DeSantis:1995}, for several values of $N$.
The solid line is the theoretical value of the
Edward-Anderson order parameter $q_{EA}(T_D)$ for $N\rightarrow\infty$
\protect\cite{DeSantis:1995}. The dashed line locates the value we use to
define the relaxation time $\tau$.}
\end{figure}

\begin{figure}[h]
\psfig{figure=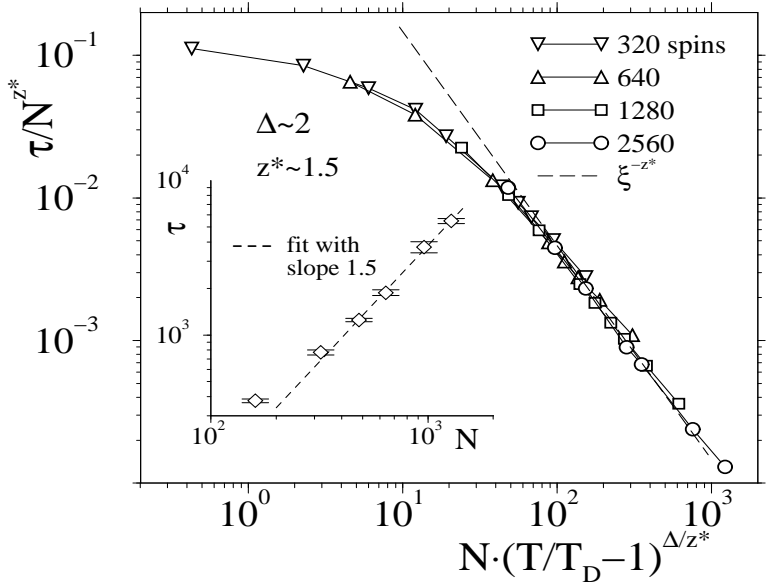,width=10.0cm,height=7.9cm}
\caption{Log-log plot of the scaled relaxation time $\tau/N^{z^*}$
vs. the scaled distance in temperature $N(T/T_D-1)^{\Delta/z^*}$ from the
dynamical transition temperature, choosing $z^*=1.5,
\Delta/z^*\approx 1.3$. The inset is a log-log plot of $\tau(T=T_D)$ vs. $N$.}
\end{figure}

\begin{figure}[h]
\psfig{figure=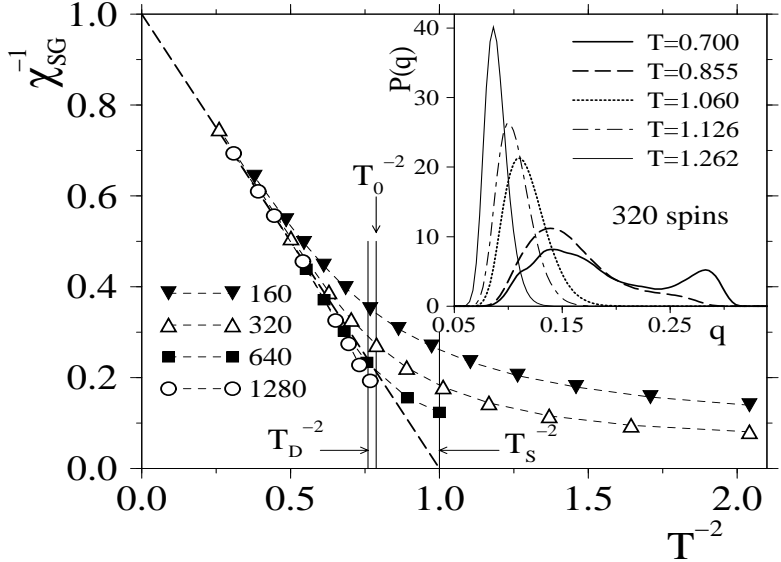,width=10.0cm,height=7.9cm}
\caption{Inverse of the spin glass susceptibility $\chi^{-1}_{SG}$
plotted vs. $T^{-2}$. Different symbols correspond to different values
of $N$. The number of different samples used is $300$ for $N=160$,
$200$ for $N=320$ and $100$ for $N=640$ and $1280$. Arrows locate
$T_s,T_0,T_D$ ~\protect\cite{DeSantis:1995}.  The inset shows the order
parameter distribution $P(q)$ vs. $q$ for $N=320$ and various
temperatures.}
\end{figure}

\end{document}